\documentclass{iopart}
\usepackage{iopams}

\usepackage[english]{babel}

\usepackage{amsfonts}
\usepackage{amssymb}
\usepackage{amsthm}
\usepackage{amscd}
\usepackage{epsfig}

\usepackage{fancyvrb}
\usepackage{afterpage}

\def\p{\partial}

\newcommand{\cU}{{\cal U}}
\newcommand{\cV}{{\cal V}}
\newcommand{\Sc}{Schr\"odinger }

\newtheorem{lemma}{Lemma}
\newtheorem{teo}{Theorem}

\newcommand{\pmm}{\partial_{(m)}}

\newcommand{\wtcU}{\widetilde{{\cal U}}}

\newcommand{\rf}[1]{(\ref{#1})}
 \def\be{\begin{equation}}
\def\ee{\end{equation}}
\def\bea{\begin{eqnarray}}
\def\eea{\end{eqnarray}}

\begin{document}
\title
{Singular matrix Darboux transformations
in the inverse scattering method}

\author{
A. A. Pecheritsin, A. M. Pupasov and
 Boris F. Samsonov}

\address{Physics Department, Tomsk State
 University, 634050 Tomsk, Russia.}

 \ead{\mailto{pecher@phys.tsu.ru, pupasov@phys.tsu.ru, samsonov@phys.tsu.ru}
 }

 \begin{abstract}
Singular Darboux transformations, in contrast to the conventional
ones, have a singular matrix as a coefficient before the
derivative. We incorporated such transformations into a chain of
conventional transformations and presented determinant formulas
for the resulting action of the chain.
A determinant representation
of the Kohlhoff-von Geramb solution to the Marchenko equation
 is given.
 \end{abstract}


\section{Introduction}

The idea to use Darboux
(or equivalently supersymmetric, or simply SUSY) transformations
for solving the inverse scattering problem for the one-dimensional
one-channel
Schr\"odinger equation was formulated in the most clear way by Sukumar
\cite{Sukumar}.
It was essentially advanced by
 introducing phase
equivalent transformations \cite{Baye1}-\cite{Baye4}
generating potentials with modified spectra.
This approach permitted to solve
the long-standing problem of the deep or shallow nature of the
nucleus-nucleus potentials \cite{Baye1}
(see also the review paper \cite{Baye-Sparenberg:JPA2004}).
In \cite{my2002} this approach was reformulated
by replacing a chain of phase equivalent
(more precisely isophase) transformations by
an equivalent $N$th-order transformation and using its Wronskian
representation based on Crum-Krein formulas \cite{Crum,Krein}.
In this way both the interaction potential and solutions of the
corresponding Schr\"odinger equation acquire a
 compact analytic form which we call the determinant representation
\cite{my2002}.
Moreover, in \cite{SS2003} the authors
 have shown how such
SUSY transformations could yield correct phase shift effective
range expansions.

The generalization of this approach to multichannel scattering confronted
with numerous difficulties.
In particular, the problem of reconstructing, by the technique of Darboux
transformations,
the neutron-proton potential proposed in
1957 by Newton and Fulton was solved only quite recently \cite{our_prepr}.
Obtained solution illustrates a general method of solving
the multichannel inverse scattering problem which may become an
alternative to the Gelfand-Levitan \cite{gelfand:51}
 or Marchenko \cite{marchenko:55} methods when applied to
the two-channel inverse scattering problem.
The main idea of the method is to divide the
problem in two steps.
This is possible since any $2\times2$ scattering matrix is uniquely
determined by its eigenphase shifts and the mixing parameter which may be
fitted to their experimental values independently.

First,
one constructs an uncoupled potential best fitting the given eigenphase
shifts.
Since the potential is supposed to be uncoupled,
this problem reduces to
 solving $n$ ($n$ is the number of channels)
``one-channel problems".
At this step, hence, one can use well elaborated ``one-channel techniques''
(for a review see e.g. \cite{Baye-Sparenberg:JPA2004}).
At the second step, using recently proposed eigenphase preserving (EPP)
transformations \cite{pupasov:10}, one constructs the final potential
by fitting the given mixing parameter only and keeping unchanged the eigenphase
shifts.

We have to emphasis that the formalism realizing one channel
transformations is essentially different of that realizing two-channel EPP
transformations.
Of course, any first order in derivative one-channel transformation may be written
in a matrix form but the
rank of the matrix coefficient before the derivative is strictly less than
$n$ so that this matrix is non-invertible.
We call such transformations {\em singular Darboux (SUSY) transformations}.
By this reason the method developed in \cite{pecheritsin:04} for replacing a
chain containing such transformations by a single $n$th order
transformation cannot be directly applied in this case.
Note that such SUSY transformations,
as far as we know, for the first time were studied in
\cite{andrianov:97}.
The main aim of the current paper is to extend the method
of paper \cite{pecheritsin:04}
  by accepting
singular transformations of a special type
as links of Darboux transformations chain.
This permits us to treat both singular and EPP transformations in a
unified  way.
All intermediate quantities,
such as solutions of the intermediate Schr\"odinger equations,
 do not enter into final
expressions for the potential and wave functions so that they are
expressed in terms of solutions of the initial equation only.

The paper is organized as follows.
In the next section we
introduce necessary notations and describe the mathematical nature of the
problem.
In particular, we emphasis the difference between  regular and singular
Darboux transformations
 and present the latter
in a form which allows us to treat both usual and singular transformations
as links with equal rights of a transformation chain.
The determinant formulas for the transformed solutions
and potential are derived.
In the third section we apply these results to construct
the simplest $2\times 2$
exactly-solvable potential
describing  $^3S_1$ $-$ $^3D_1$ neutron-proton elastic collisions.
In Conclusion
we discuss main results and outline a possible line of future investigations.

\section{Chains of Darboux transformations with singular links}

\subsection{Preliminaries}

We start with the matrix \Sc equation
\begin{equation} \label{matrix_Schr}
\fl
H_0 \Psi_E(x) = E \Psi_E(x)\,,\quad H_0 = - I_n \p^2 +
V_0(x)\,,\quad \p:=\frac{\p}{\p x}\,.
\end{equation}
where $\Psi_E(x) = \bigl(\psi_{1E}(x), \ldots \psi_{nE}(x)\bigr)^t$ is a
vector-valued function, $I_n$ is $n\times n$ identity matrix,
$V_0(x)$ is an $n\times n$ real and symmetric (potential) matrix.
In general, $x$ is a real variable which may belong to the whole real axis,
to a semiaxis or to a closed interval of the real axis.
In this section we do not associate equation \rf{matrix_Schr} with any
spectral problem and consider it as a system of differential equations of
a special type. In the next section we shell treat it as
a radial matrix \Sc equation.

First, to fix notations, we would like to remind main notions
about matrix Darboux transformations and their chains
\cite{pecheritsin:04}.


 Suppose that we know $N$ matrix solutions to equation (\ref{matrix_Schr})
 corresponding to different eigenvalue matrices
 $\Lambda_k\ne  \Lambda_l$ for $k\ne l$,
 \[
 H_0\cU_k=\cU_k\Lambda _k\,,\quad k=1,\ldots ,N\,.
 \]
Here, in general, the spectral parameter $\Lambda_k$
may be an arbitrary constant matrix but, following paper \cite{pecheritsin:04},
we assume that it is a diagonal matrix with real entries.
Moreover, for the case of equal thresholds, that we bear in mind,
$\Lambda_k$ is a number.

 For the first transformation step we
 take matrix $\cU_1$
(we call it the {\em transformation matrix})
 and construct the transformation
 operator
 \begin{equation} \label{L1}
 L_{1 \leftarrow 0} = I_n\p  - F_1\,,\quad F_1=(\p\cU_1)\cU_1^{-1}\,.
 \end{equation}
 Note that it can be applied not only on vector-valued
 functions like $\Psi_E$ but also on matrix-valued like $\cU_{\,2}$,
 \ldots , $\cU_N$.
 In this way we get the matrix solutions
 $\cV_2=  L_{1 \leftarrow 0}\cU_{\,2}$, \ldots , $\cV_N=  L_{1 \leftarrow 0}\cU_N$
 of  the equation with the potential
\[
 V_1=V_0-2\p F_1\,.
\]
 Now $\cV_2$ can be taken
 as transformation matrix for the Hamiltonian $H_1=-I_n\p^2+V_1$
 to produce the potential
 \[
  V_2=V_1-2\p\bigl((\p\cV_2)\cV_2^{-1}\bigr)=V_0-2\p F_2\,,\quad
  F_2=F_1+(\p\cV_2)\cV_2^{-1}
  \]
 and the transformation operator
 $L_{2 \leftarrow 1} = I_n\p  - (\p\cV_2)\cV_2^{-1}$
 and so on, till one gets the potential
 \[
 V_N=V_0-2\p F_N
  \]
 with $F_N$ defined recursively
 \[
  F_N=F_{N-1}+(\p Y_N)Y_N^{-1}\,,\quad N=1,2,\ldots \ \,,\quad F_0=0
 \]
 and $Y_N$ being a matrix-valued solution
 to the \Sc equation at $(N-1)$th step of
 transformations,
\[
H_{N-1}Y_N=Y_N\Lambda_N\,,\quad H_{N-1}=-\p^2+V_{N-1}\,.
\]
The matrix $Y_N$ is obtained by the action of the chain of $N-1$
transformations applied to the matrix $\cU_n$,
 \[
 Y_N = L_{(N-1) \leftarrow (N-2)}\ldots L_{2 \leftarrow 1}L_{1 \leftarrow 0} \cU_N
 \equiv  L_{(N-1) \leftarrow 0} \cU_N
 \]
and it produces  operator
 $L_{N \leftarrow (N-1)}=-I_n\p+(\p Y_N)Y_N^{-1}$
of the final transformation step
 for the chain of
 $N$ transformations.

 To get in this way the final potential $V_N$
resulting from the chain of $N$ transformations,
 one has to
 calculate all intermediate transformation matrices $Y_j$,
 $j=2,\ldots,N$ performing
 a huge amount of unnecessary work even for the one-channel case.
 In practical calculations one is able to perform only few steps
 which restricts considerably possible applications of the method.
 Fortunately, for the one-channel case there exists what that are
 called Crum-Krein \cite{Crum,Krein}  determinant formulas.
 Their multichannel generalization is
 given in \cite{pecheritsin:04}.
These formulas allow one to omit
 all intermediate steps and go from $H_0$ directly to $H_N$.
 This is achieved by expressing the
$N$th order transformation operator
  \[
  L_{N\leftarrow0}=
 L_{N \leftarrow (N-1)}\ldots L_{2 \leftarrow 1}L_{1 \leftarrow 0}
 \]
 realizing the
 resulting action of the chain
in terms of solutions of the initial \Sc equation \rf{matrix_Schr} only.

We would like to emphasis that the authors of \cite{pecheritsin:04}
 considered
chains of first order transformation operators of the form \rf{L1} where
the coefficient before the derivative is a regular matrix which, being
$x$-independent, can always be reduced to identity matrix.
Nevertheless, as
it was first stressed by Andrianov et al \cite{andrianov:97},
one of the main features of matrix transformations is that the
coefficient before the derivative may be a singular matrix.
The method developed in \cite{pecheritsin:04} is not directly applicable to
this case and needs some modifications.
Below
we give necessary modifications of that method for a particular type of singular
matrices that appear when the SUSY method is used for solving
the inverse scattering problem
\cite{our_prepr}.

\subsection{Singular matrix Darboux transformations of a special type}
Assume  that
the initial quantum system consists of two non-interacting subsystems $(I)$
and $(II)$ so that
the matrix $V_0(x)$
is block-diagonal
\[
V_0(x) =\left(
\begin{array}{cc}
V_0^{(I)}(x) & 0 \\ 0 & V_0^{(II)}(x)
\end{array}
   \right).
\]
Here $V_0^{(I)}$ is an $m\times m$ matrix, $m=1,\ldots,n-1$,
and $V_0^{(II)}$ is an
$(n-m)\times(n-m)$ matrix.
Since the total potential matrix $V_0(x)$ is assumed to be real and
symmetric, this implies that sub-matrices $V_0^{(I)}$ and $V_0^{(II)}$ are
also real and symmetric.
In this case, equation (\ref{matrix_Schr})
splits into two independent matrix equations
\begin{eqnarray}
h_0^{(I)}\Psi_E^{(I)}(x) & = & E \Psi_E^{(I)}(x)\,,\quad
h_0^{(I)}=- I_m \p^2 + V_0^{(I)}(x)
\label{shred-I} \\
h_0^{(II)}\Psi_E^{(II)}(x) & = & E
\Psi_E^{(II)}(x)\,,\quad
h_0^{(II)}=- I_{n-m} \p^2 + V_0^{(II)}(x)\,,
\nonumber 
\end{eqnarray}
where $\Psi_E^{(I)}(x) = \bigl(\psi_{1E}(x), \ldots,\psi_{mE}(x)\bigr)^t$,
$\Psi_E^{(II)}(x) = \bigl(\psi_{m+1E}(x), \ldots,\psi_{nE}(x)\bigr)^t$.

Assume that we want to realize the first order transformation over a subsystem
only, say for definiteness over the subsystem (I). Evidently this should not affect the
subsystem (II).
Therefore the corresponding Darboux transformation operator,
which we denote $L^{(m)}$, has the form
\begin{equation}\label{darbu-m}
L^{(m)} = \left(
\begin{array}{cc} I_m & 0 \\ 0  & 0 \end{array}\right) \p +
\left(
\begin{array}{cc} - (\p{\widetilde{\cU}} ) \wtcU^{-1} & 0 \\ 0 & I_{n-m} \end{array}\right)
\end{equation}
where the matrix $\wtcU$ is an eigensolution of the Hamiltonian
$h_0^{(I)}$
\[
h_0^{(I)} \wtcU = \wtcU \tilde\Lambda\,.
\]

We would like to emphasis that the coefficient before the derivative in
\rf{darbu-m} is a singular matrix.
Therefore it can never be
presented in the form $I_n\p+w$ and such transformations cannot be
directly incorporated into the usual chain of matrix Darboux transformations
as considered in \cite{pecheritsin:04}.
Nevertheless, theorems proven in \cite{pecheritsin:04} have a more general
character than Darboux transformation of the matrix Schr\"odinger equation.
Actually, they represent a closure of a special recursion procedure.
Below we show that with short modifications they may be applied
to the
current case also but first we will rewrite the operator \rf{darbu-m} in
a form more suitable for our purpose.

First we note that for any matrix $\cU$ of the form
\begin{equation}\label{cU}
\cU = \left(
\begin{array}{cc} \wtcU & 0 \\ 0 & I_{n-m}
\end{array}\right).
\end{equation}
the following property holds
\[
(\p{\cU}) \cU^{-1}  =
\left(
\begin{array}{cc} (\p{\wtcU}) \wtcU^{-1} & 0 \\ 0 & 0
\end{array}
\right).
\]
Next, if we move the matrix $I_{n-m}$ from the second term
in the right hand side of equation \rf{darbu-m} to its first term,
the operator $L^{(m)}$ takes the form
\begin{equation} \label{darbu-pm}
L^{(m)} = D_m - (\p{\cU}) \cU^{-1}.
\end{equation}
Here the matrix valued operator
\begin{equation}\label{Dm}
D_m=
\left(
\begin{array}{cc} I_m\p & 0 \\ 0  & I_{n-m} \end{array}\right)
\end{equation}
differentiates the first $m$ entries
of a vector
$\Phi=(\varphi_1,\ldots,\varphi_m,\varphi_{m+1},\ldots,\varphi_n)$
and keeps unchanged any its component $\varphi_l$ for $l>m$.
Similarly, while acting on an $n\times n$ matrix, it differentiates the
first $m$ entries of each column of this matrix and keeps unchanged any
other entry. Below we will also use the composition of operators \rf{Dm},
$D_m^M=\underbrace{D_m\cdot\ldots\cdot D_m}_{\mbox{\tiny M times}}$.
When we need either differentiate or keep unchanged a separate component
$\varphi_l$, $1\le l\le n$ of vector
$\Phi$, we will use the notation $\pmm$ defined as
\begin{equation}\label{pm}
\pmm \varphi_l =\left\{
\begin{array}{ll}
\p \varphi_l, & l\le m\,, \\
\varphi_l ,   & l > m\,.
\end{array}    \right.
\end{equation}
Note that the value of $m$ is fixed by the dimension of the subsystem
$(I)$.

Expression \rf{darbu-pm} has the same structure as the usual Darboux
transformation where
$I_n\p$ is replaced by $D_m$.
It should also be
noted that the matrix $\cU$ is not solution to the initial \Sc
equation (\ref{matrix_Schr}) but
it is constructed from solutions to
the \Sc equation for the subsystem $(I)$ (\ref{shred-I}).
In the next section, we will present a generalization of the scheme
 developed in \cite{pecheritsin:04} which permits us to incorporate
 transformations such as the one given in \rf{darbu-pm} into a chain of usual
 transformations but first we need to introduce some new notations.

\subsection{Notations}
Let $\cU_k$, $k = 1, \ldots,N$ be a collection of $n\times n$ matrices
\[
\cU_k = \left(
\begin{array}{llll}
u_{1,1;k} & u_{1,2;k} & \ldots & u_{1,n;k} \\
u_{2,1;k} & u_{2,2;k} & \ldots & u_{2,n;k} \\
\ldots &\ldots &\ldots &\ldots \\
u_{n,1;k} & u_{n,2;k} & \ldots & u_{n,n;k}
\end{array}
\right).
\]
Any matrix $\cU_k$ may be presented
as either a collection of $n$--dimensional column-vectors
$U_{j;k}=(u_{1,j;k}, \ldots ,u_{n,j;k})^t$,
$\cU_k=(U_{1;k},\ldots,U_{n;k})$, $k = 1, \ldots,N$, or a
collection of $n$--dimensional row-vectors
 $U^j_k=(u_{j,1;k}, \ldots ,u_{j,n;k})$, $j = 1,\ldots,n$,
 $\cU_k=(U_k^1,\ldots, U_k^n)^t$.

Using these matrices,
we define $nM \times nM$ matrix  $W$
\begin{equation}\label{WM-def}
\fl
W  \equiv W(\cU_1,\cU_2,\ldots, \cU_M)
  = \left(
 \begin{array}{rrrrr}
 \cU_1 & \cU_2 & \cU_3 & \ldots & \cU_M  \\
 \p\cU_1 & D_m\cU_2 & D_m\cU_3& \ldots & D_m \cU_M  \\
  \p^2\cU_1 & \p^2\cU_2 & D_m^2\cU_3& \ldots & D_m ^2\cU_M  \\
 \ldots  &  \ldots &  \ldots &  \ldots   &  \ldots \\
 \p^{M-1}\cU_1 & \p^{M-1}\cU_2 & \p^{M-1}\cU_3 & \ldots & D_m^{M-1} \cU_M
 \end{array}
 \right).
\end{equation}
This matrix is used as the left upper block in a larger $nN \times nN$, $N>M$,
 matrix
\begin{equation}\label{W-def}
\fl \setlength{\arraycolsep}{2pt}
\begin{array}{ll}
 & W(\cU_1,  \ldots,\cU_M; \cU_{M+1}, \ldots \cU_N) \\
 &  = \left(
 \begin{array}{rrrrrrr}
          &  &  &  & \cU_{M+1} & \ldots & \cU_N \\
\multicolumn{4}{c}{W(\cU_1,\cU_2,\ldots, \cU_M)}  & D_m \cU_{M+1} & \ldots &
D_m \cU_N \\
                 &  &  &  &  \ldots &  \ldots &  \ldots\\
                 &  &  &  & D_m^{M-1} \cU_{M+1} & \ldots & D_m^{M-1} \cU_N \\
\p^{M}\cU_1 & \p^{M}\cU_2 & \ldots &
\p^{M} \cU_M & D_m^M \cU_{M+1} & \ldots & D_m^M \cU_N \\
\p^{M+1}\cU_1 & \p^{M+1}\cU_2 & \ldots & \p^{M+1} \cU_M & \p D_m^M \cU_{M+1} &
 \ldots & \p D_m^M \cU_N \\
\p^{M+2}\cU_1 & \p^{M+2}\cU_2 & \ldots &
\p^{M+2} \cU_M & \p^2D_m^M \cU_{M+1} &
 \ldots & \p^2D_m^M \cU_N \\
\ldots  &  \ldots &  \ldots &  \ldots  &  \ldots &  \ldots &  \ldots\\
\p^{N-1}\cU_1 & \p^{N-1}\cU_2 & \ldots & \p^{N-1} \cU_M &
\p^{N-M-1}D_m^M \cU_{M+1} & \ldots & \p^{N-M-1}D_m^M \cU_N
 \end{array}
 \right).
\end{array}
\end{equation}
Note that the first $M$ matrices $\cU_k$, $k=1,\ldots,M$  in \rf{W-def}
have the structure fixed by equation (\ref{cU})
whereas the structure of the matrices
 $\cU_k$ for $k=M+1,\ldots,N$ is not fixed.
For $M=N$ all the matrices in this formula have the special structure (\ref{cU})
and the matrix \rf{W-def} reduces to \rf{WM-def}.

Below we will also need a set of $(nN+1)\times(nN+1)$ matrices $W_j$,
$j=1,\ldots,n$
obtained from (\ref{W-def})
by adding to the right a column constructed with the help of $n$-vector
$\Psi=(\psi_1,\ldots,\psi_n)$, to the bottom a row constructed from the
$j$th row $U^j_k$ of matrices $\cU_k$ and the right bottom corner is
filled with a derivative of $j$th element of the vector $\Psi$,
\begin{equation} \label{Wj-def}
\fl \setlength{\arraycolsep}{2pt}
\begin{array}{ll}
& W_j(\cU_1, \ldots,\cU_M; \cU_{M+1}, \ldots, \cU_{N}, \Psi) = \\
& = \left(
 \begin{array}{rrrrrrr}
  \multicolumn{6}{c}{W(\cU_1, \ldots,\cU_M; \cU_{M+1}, \ldots, \cU_{N})} &
    \begin{array}{r}
  \Psi \\ \ldots \\ D_m^{M-1}\Psi \\ D_m^M \Psi \\ \p D_m^M \Psi \\ \ldots \\
\p^{N-M-1}D_m^M \Psi
    \end{array}
  \\
 \p^NU_1^{j} & \ldots & \p^NU_M^{j} & \p^{N-M}D_m^MU_{M+1}^{j} & \ldots
& \p^{N-M}D_m^MU_N^{j}
     &
   \p^{N-M}\pmm^M\psi_j
 \end{array}
  \right).
\end{array}
\end{equation}
Finally we introduce matrices
$W_{i,j}(\cU_1, \ldots, \cU_M; \cU_{M+1},\ldots, \cU_N)$, $i,j=1,\ldots,n$,
 obtained from the matrix
(\ref{W-def}) by replacing in its last matrix row
the matrices
$\p^{N-M-1}D_m^M\cU_k$ (or $D_m^{M-1}\cU_k$ in the case  $N=M$)
by matrices
$\cU_k^{ij}$, $k=1,\ldots,N$,
\begin{equation}\label{Wij-def}
\fl \setlength{\arraycolsep}{2pt}
\begin{array}{ll}
& W_{i,j}(\cU_1,  \ldots,\cU_M; \cU_{M+1}, \ldots, \cU_N) = \\
 & = \left(
 \begin{array}{rrrrrrr}
\multicolumn{4}{c}{W(\cU_1,\cU_2,\ldots, \cU_M)} & \cU_{M+1} & \ldots & \cU_N \\
                 &  &  &  & D_m \cU_{M+1} & \ldots & D_m \cU_N \\
                 &  &  &  &  \ldots &  \ldots &  \ldots\\
                 &  &  &  & D_m^{M-1} \cU_{M+1} & \ldots & D_m^{M-1} \cU_N \\
\p^{M}\cU_1 & \p^{M}\cU_2 & \ldots &
\p^{M} \cU_M & D_m^M \p\cU_{M+1} & \ldots & D_m^M \cU_N \\
\p^{M+1}\cU_1 & \p^{M+1}\cU_2 & \ldots &
\p^{M+1} \cU_M & \p D_m^M \cU_{M+1} & \ldots & \p D_m^M \cU_N \\
\ldots  &  \ldots &  \ldots &  \ldots  &  \ldots &  \ldots &  \ldots\\
\p^{N-2}\cU_1 & \p^{N-2}\cU_2 & \ldots &
\p^{N-2} \cU_M & \p^{N-M-2}D_m^M \cU_{M+1} & \ldots & \p^{N-M-2}D_m^M \cU_N \\
\cU_1^{i,j} & \cU_2^{i,j} & \ldots & \cU_M^{i,j} & \cU_{M+1}^{i,j}
& \ldots & \cU_N^{i,j}
 \end{array}
 \right).
\end{array}
\end{equation}
The matrices $\cU_k^{i,j}$, $i,j=1,\ldots,n$, $k=1,\ldots,N$, are
constructed from the matrix $\p^{N-M-1}D_m^M\cU_k$ (or
$D_m^{M-1}\cU_k$ when $M=N$) by replacing its $j$th row with the
$i$th row of the matrix $\p^{N-M}D_m^M\cU_1$ (or $\p
D_m^{M-1}\cU_k$ when $M=N$).

Before proving our main result we need two lemmas.

\subsection{Two lemmas}

Consider the matrix
 \[
 A=\left(
 \begin{array}{cccccc}
 a_{1,1} & \ldots & a_{1,p} & a_{1,p+1} & \ldots & a_{1,p+n}\\
        &  \ldots &       &          & \ldots &         \\
 a_{p,1} & \ldots & a_{p,p} & a_{p,p+1} & \ldots & a_{p,p+n}\\
 b_{1,1} & \ldots & b_{1,p} & b_{1,p+1} & \ldots & b_{1,p+n}\\
 b_{2,1} & \ldots & b_{2,p} & b_{2,p+1} & \ldots & b_{2,p+n}
 \end{array}
 \right).
\]
 Let $a$ be the $p\times p$ submatrix of $A$
 with the entries $a_{i,j}$, $i,j=1,\ldots ,p$.
 Denote $m_{jk}$ the minor of $A$ embordering $a$ with
 $j$th ($j=1,2$) row composed of  $b_{j,i}$, $j=1,2$,
 $i=1,\ldots ,p$ and $k$th column ($p<k\le p+n$).
 (For the definition of the embordering minor see Appendix.)
 Let also $m_{jk}^{ts}$ be the minor obtained from  $m_{jk}$
 by replacing its $s$th row composed of
  $a_{s,j}$, ($s\le p$) with $(p+t)$th row composed of
   $b_{t,i}$ ($t=1,2$).
   Let now $a^{ts}$ be obtained from $a$ with the help of the same
   replacement, i.e. with the replacement of its $s$th row
   composed of $a_{s,j}$, ($s\le p$) by $(p+t)$th row of $A$ composed of
    $b_{t,j}$ ($t=1,2$).

  \begin{lemma}{\rm\cite{pecheritsin:04}}\label{lm:minors}
 If  $|a|\ne 0$ then
 the following determinant identity takes place
\[
 |a|\, m_{jk}^{ts} = |a^{ts}| m_{jk} - |a^{js}| m_{tk}\,.
\]
\end{lemma}

The second lemma establishes a rule for differentiating the ratio of two
determinants.
\begin{lemma}\label{phij-dif}
Let
\begin{equation} \label{phi-j}
 \varphi_j =
 \frac{|W_j(\cU_1, \ldots, \cU_M; \cU_{M+1}, \ldots, \cU_N, \Psi)|}
{|W(\cU_1, \ldots, \cU_M; \cU_{M+1}, \ldots, \cU_N)|}\,.
\end{equation}
Then
\[
 \p \varphi_j =
 \frac{\Gamma_j - \sum\limits_{l=1}^n \varphi_l |W_{j,l}(\cU_1,
\ldots, \cU_M; \cU_{M+1}, \ldots, \cU_N)|}%
{|W(\cU_1, \ldots, \cU_M; \cU_{M+1}, \ldots, \cU_N)|}
\]
where $\Gamma_j$ is the determinant of the matrix obtained from
$W_j(\cU_1, \ldots, \cU_M; \cU_{M+1}, \ldots, \cU_N, \Psi)$
by differentiating its last row.
\end{lemma}

\begin{proof}
In the expression for the derivative of the fraction \rf{phi-j}
\begin{eqnarray}
 \p \varphi_j =
 - \frac{\p |W(\cU_1, \ldots, \cU_M; \cU_{M+1}, \ldots, \cU_N)|}
{|W(\cU_1, \ldots, \cU_M; \cU_{M+1}, \ldots, \cU_N)|} \varphi_j
\nonumber
\\
+
\frac{\p |W_j(\cU_1, \ldots, \cU_M; \cU_{M+1}, \ldots, \cU_N,
\Psi)|} {|W(\cU_1, \ldots, \cU_M; \cU_{M+1}, \ldots, \cU_N)|}
 \label{dphi_j-1}
\end{eqnarray}
we  first analyze the derivative of the determinant
$|W(\cU_1, \ldots, \cU_M; \cU_{M+1}, \ldots, \cU_N)|$.
Taking into account the structure of matrix
(\ref{W-def}), one can see that a non zero
contribution to this derivative gives a term
corresponding to the differentiation of
the last matrix row only
\begin{equation}
\fl
\p |W(\cU_1, \ldots, \cU_M; \cU_{M+1}, \ldots, \cU_N)| =
\sum\limits_{l=1}^n |W_{l,l}(\cU_1, \ldots, \cU_M; \cU_{M+1}\,,
\ldots, \cU_N)|.
 \label{dW}
\end{equation}
Note, that since $W_{l,l} = 0$ for $l>m$, then,
 in the case $M=N$, the summation in this expression goes up to $l=m$ only.

Similarly, for the derivative of $W_j$, one gets
\begin{equation}
\p |W_j(\cU_1, \ldots, \cU_M; \cU_{M+1}, \ldots, \cU_N, \Psi)| =
\Gamma_j + \sum\limits_{l\ne j}^n \Delta_j^l\,.
 \label{dW_j}
\end{equation}
Here $\Delta_j^l$ is the determinant obtained from $W_j$
by differentiating the  $l$th row in matrices of the next to last
 matrix row of $W_j$.

We will use Lemma \ref{lm:minors} to calculate $\Delta_j^l$.
As matrix $a$ we choose $W$.
By adding to $W$ two rows and one column,
 we obtain matrix $A$ from
Lemma \ref{lm:minors}.
One row and the column are exactly the last row and the last column
of the matrix $W_j$ \rf{Wj-def}.
The second added row is the $l$th row of matrices
$\p^{N-M+1}D_m^{M-1}\cU_k$, $k=1,\ldots,N$, i.e.
$\p^{N-M+1}D_m^{M-1}U_k^l$, $k=1,\ldots,N$.
Finally, to obtain matrix $A$ we fill the right bottom corner
 with the element $\p^{N-M+1}\pmm^{M-1}\psi_l$.
Thus, identifying in Lemma \ref{lm:minors}
$m^{ts}_{jk}$ with $\Delta^l_j$, $a^{ts}$ with $W_{ll}$, $m_{jk}$ with
$|W_j|$, $a^{js}$ with $W_{j,l}$ and $m_{tk}$ with $|W_l|$, we find
\begin{eqnarray}
\Delta_j^l =  \frac{|W_{l,l}(\cU_1, \ldots, \cU_M; \cU_{M+1}, \ldots, \cU_N)|\, |W_j|}%
{|W(\cU_1, \ldots, \cU_M; \cU_{M+1}, \ldots, \cU_N)|}
\nonumber
\\
-\frac{ |W_{j,l}(\cU_1, \ldots, \cU_M; \cU_{M+1},\ldots, \cU_N)|\, |W_l|}
{|W(\cU_1, \ldots, \cU_M; \cU_{M+1}, \ldots, \cU_N)|}\,.
\nonumber
\end{eqnarray}
From here and (\ref{dW_j}) it follows that
\begin{eqnarray}
\fl
\p |W_j(\cU_1, \ldots, \cU_M; \cU_{M+1}, \ldots, \cU_N, \Psi)| =
\Gamma_j
\nonumber
 \\
 \fl
 \!\! +\!\! \sum\limits_{l\ne j}^n \big[ |W_{l,l}(\cU_1, \ldots,
 \cU_M; \cU_{M+1}, \ldots, \cU_N)| \varphi_j \! - \!
 |W_{j,l}(\cU_1, \ldots, \cU_M; \cU_{M+1}, \ldots, \cU_N)|
\varphi_l \big]\,.
\nonumber
\end{eqnarray}
We complete the proof by substituting (\ref{dW}) and (\ref{Wj-def})
into  (\ref{dphi_j-1}),
\end{proof}

Note that this lemma plays a crucial role in the proof of our main results
since it makes here applicable
 with only short modifications
the proof of similar theorems given in
\cite{pecheritsin:04}.

\subsection{Transformation of a vector}
Let us consider a chain of  $N$ matrix
first order Darboux transformations, where the first $M$
transformations are singular (i.e., transformations of the subsystem $(I)$)
and the remaining $N-M$ transformations are usual matrix transformations.
 We will denote any particular transformation in this chain
as $L_{j \leftarrow (j-1)}$, $j=1,\ldots,N$ possibly with the superscript
$(m)$ if this is a singular transformation,
$L_{j \leftarrow (j-1)}^{(m)}$,
and by $L_{N \leftarrow 0} $ will be denoted the superposition of $N$
first order transformations so that
\be\label{LN0}
L_{N \leftarrow 0} = L_{N \leftarrow (N-1)} \ldots L_{(M+1)
\leftarrow M}
L^{(m)}_{M \leftarrow (M-1)} \ldots L^{(m)}_{1 \leftarrow 0}
\ee
where
\be\label{Lk_k-1}
 L_{k \leftarrow (k-1)} = \p - (\p Y_k) Y_k^{-1},\quad
k=M+1,\ldots,N
\ee
and
\be\label{Lmk_k-1}
L^{(m)}_{k \leftarrow (k-1)} = D_m - (\p Y_k) Y_k^{-1},\quad
k=1,\ldots M
\ee
with
\be\label{Y_k-def}
Y_1 = 0, \hspace{1cm} Y_k = L_{k \leftarrow 0}
\cU_k,\hspace{0.5cm} k = 2, \ldots,N\,.
\ee
The resulting action  of the chain is a transformation from the initial
potential $V_0$ to the final potential $V_N$.

In the case $M=N$ there are only singular transformations.

We also note that the operator $L_{N \leftarrow 0}$ can be applied not
only on solutions of the \Sc equation but also on any vector-valued
function $\Psi=(\psi_1,\ldots,\psi_n)^t$.

\begin{teo}\label{teo:mShL_N}
Chain of $M$ singular and $N-M$ ordinary Darboux transformations acts on
a vector $\Psi=(\psi_1,\ldots,\psi_n)^t$
as follows
\[
\Phi =L_{N \leftarrow 0} \Psi =
(\varphi_{1},\ldots,\varphi_{n})^t
\]
where
\[
\varphi_{j}= \frac{|W_j(\cU_1,\ldots,\cU_M; \cU_{M+1}, \ldots,
\cU_N,\Psi)|} {|W(\cU_1, \ldots,\cU_M; \cU_{M+1}, \ldots,
\cU_N)|}
\]
with  $W_j(\cU_1,\ldots,\cU_M; \cU_{M+1}, \ldots,\cU_N,\cU_N,\Psi)$
 and
$W(\cU_1,\ldots,\cU_M; \cU_{M+1}, \ldots, \cU_N,\cU_N)$
defined in (\ref{Wj-def}) and (\ref{W-def}) respectively.
\end{teo}

\begin{proof}

We only give main ideas of the proof,
since, at it was already mentioned,
after proving Lemma \ref{phij-dif}, the proof of the similar theorem given
in \cite{pecheritsin:04} is applicable here with short modifications.

Following the method of paper \cite{pecheritsin:04}, we will use
the perfect induction method for proving the theorem.
There are here two discrete variables, $M$ and $N$.
Therefore first we will prove the statement for $M=N$ and then for a fixed
value of $M$, we will comment on the case $N>M$.

For $M=N=1$
 there is only one singular transformation
\begin{equation} \label{L_1_Psi}
\Phi =L^{(m)}_{1 \leftarrow 0} \Psi = D_m \Psi - (\p \cU_1)
\cU_1^{-1} \Psi\,.
\end{equation}
Following the same lines as in \cite{pecheritsin:04}, one finds
\begin{equation} \label{phi_l}
\varphi_{l} = \frac{|W_l(\cU_1,\Psi)|}{|\cU_1|}
\end{equation}
where
\[
W_l(\cU_1,\Psi) =
\left( \begin{array}{cr} \cU_1 & \begin{array}{r} \psi_{1} \\
\ldots \\ \psi_{n}
          \end{array} \\
\begin{array}{rcr} \p u_{l,1;1} & \ldots &\p  u_{l,n;1}  \end{array}
 & \pmm \psi_{l}
\end{array} \right)
\]
meaning that
the theorem holds for $M=N=1$.

For $M=N$ assume the theorem to hold for the chain of $N-1$ singular
transformations, i.e.
\begin{eqnarray}
& \Theta = L_{(N-1) \leftarrow 0}
\Psi_E = (\theta_1, \ldots,\theta_n)^t\,,
\label{L_N-1_Psi}\\
& \theta_j =\displaystyle{ \frac{|W_j(\cU_1,\ldots,\cU_{N-1},\Psi)|}%
{|W(\cU_1,\ldots,\cU_{N-1})|}}\,, \label{theta_j}
\end{eqnarray}
and  prove it for the chain of $N$
 transformations.

Operator $L_{N\leftarrow (N-1)}$ transforms the vector $\Theta$
\rf{L_N-1_Psi}
into the vector $\Phi =L_{N\leftarrow (N-1)}\Theta$.
To find the result of this transformation,
i.e. the operator
$L_{N\leftarrow0}=L_{N\leftarrow(N-1)}L_{(N-1)\leftarrow0}$ \rf{LN0}
it is necessary according to \rf{Lk_k-1}
to calculate matrix $Y_N=L_{(N-1)\leftarrow0}\cU_N$ which
 is a collection of columns
$Y_N=(Y_{1;N},\ldots,Y_{n;N})$.
According to (\ref{Y_k-def}),
each column transforms as an $n$-vector with the result given in
 (\ref{theta_j})
 where $\Psi $ should be replaced by a column $U_{i;N}$
  of the matrix  $\cU_N=(U_{1;N}\ldots U_{n;N})^t$.
Therefore for the entries $y_{ji}$ of the vector
$Y_{j;N}=(y_{j1},\ldots y_{jn})^t$, one gets
\[
y_{ji}= \frac{|W_j(\cU_1,\ldots,\cU_{N-1},U_{i;N})|}%
 {|W(\cU_1, \ldots,\cU_{N-1})|}\,.
\]

The determinant of the matrix  $Y_N=(y_{ji})$
can be calculated with the aid of the
 Sylvester (see Appendix) identity
 \begin{equation}
 |Y_N|=\frac{|W(\cU_1, \ldots,\cU_{N})|}%
{|W(\cU_1, \ldots, \cU_{N-1})|}\,.
\label{detY_N}
\end{equation}
According to (\ref{Lmk_k-1}),
(\ref{L_1_Psi}) and (\ref{phi_l}), we can write down the components $\varphi_{j}$
of the vector $\Phi $
as
\[
\varphi_{j} = \frac{|Y_N^j|}{|Y_N|},
\]
where
\begin{equation}\label{Y_N^j}
Y_N^j = \left(
\begin{array}{rrrr}
& & &  \theta_1\\
\multicolumn{3}{c}{Y_N} &  \ldots\\
& & & \theta_n \\
\p y_{j1} & \ldots & \p y_{jn} & \pmm \theta_j
\end{array}
\right).
\end{equation}
Note that in this case, the structure of matrix
$Y_N$ coincides with the one given in (\ref{cU}),
i.e., $y_{ji} = \delta_{ji}$ when $i>m$ or $j>m$.
Therefore the last row of matrix \rf{Y_N^j} is zero
 for $j>m$
except possibly for its right bottom corner and
the first $m$ elements  when $j\le m$.
By this reason, it remains to calculate the non-zero entries of this row
only and we will use Lemma~\ref{phij-dif} for that.

Applying Lemma~\ref{phij-dif} to  $\p y_{ji}$ yields
\begin{equation} \label{dy_ji}
 \p y_{ji} = \frac{1}{|W(\cU_1, \ldots, \cU_{N-1})|}
\bigg[\Gamma_{ji} - \sum\limits_{l=1}^n y_{li} |W_{j,l}(\cU_1,
\ldots, \cU_{N-1})|\bigg]\,.
\end{equation}
Here $\Gamma_{ji}$ is the determinant of matrix
$W_j(\cU_1,\ldots,\cU_M; \cU_{M+1}, \ldots, \cU_{N-1},U_{i;N})$
in which the elements of the last
  row are differentiated.
Quite similarly,
since for  $1 \le j \le m$ the special operator \rf{pm} acts
as the derivative, $\pmm \theta_j = \p \theta_j$,
applying Lemma~\ref{phij-dif} to \rf{theta_j}
yields
\begin{equation} \label{dtheta_j}
\fl
 \pmm \theta_j = \p \theta_j = \frac{1}{|W(\cU_1, \ldots,  \cU_{N-1})|}
\bigg[\Gamma_j - \sum\limits_{l=1}^n \theta_l
|W_{j,l}(\cU_1, \ldots, \cU_{N-1})|\bigg].
\end{equation}
Here
$\Gamma_j$ is the determinant of the matrix obtained from
$W_j(\cU_1,\ldots, \cU_{N-1},\Psi )$ by differentiating its last row.

From equations \rf{dy_ji} and \rf{dtheta_j} it follows that
 for  $1 \le j \le m$ the determinant of matrix $Y_N^j$
equals the sum of two determinants.
Furthermore, since the second terms of these formulas represent one and the
same linear combination of $y_{li}$ in \rf{dy_ji} and $\theta_l$ in
\rf{dtheta_j}, which in their turn are elements of previous rows,
the last row of the second determinant is a linear combination of the
previous rows with the coefficients $|W_{j,\cdot}|$.
By this reason this determinant vanishes.
The matrix of the first determinant consists of minors embordering
 the  block
 $W(\cU_1,\ldots, \cU_{N-1})$ in matrix $W_j(\cU_1,\ldots, \cU_N,
\Psi )$.
Applying
 the Sylvester identity to this determinant, we obtain
\begin{equation} \label{detY_N^j}
|Y_N^j| = \frac{|W_j(\cU_1,\ldots, \cU_N, \Psi )|}
{|W(\cU_1,\ldots, \cU_{N-1})|}\,.
\end{equation}
When $j > m$, $D_m \theta_j = \theta_j$ and, $\p y_{ji} = 0$
because of the structure \rf{cU} of matrix $Y_N$.
Therefore all elements of
the last row of the matrix $Y^j_N$ vanish except for the last one
which is $\theta_j$, and
 applying directly the Sylvester identity
 for calculating  $|Y_N^j|$,
one gets the same result (\ref{detY_N^j}).

Thus, the determinant $Y_N^j$ is given by (\ref{detY_N^j})
 for all $j=1,\ldots,n$.

Substituting this expression into
(\ref{Y_N^j}) and taking into account (\ref{detY_N}), we get
\begin{equation}\label{phi_j_le_m}
\varphi_{j} = \frac{|W_j(\cU_1,\ldots, \cU_N, \Psi )|}
{|W(\cU_1,\ldots, \cU_N)|}\,,
\end{equation}
thus finishing the proof of the theorem for $M=N$.

For $N=M,M+1,\ldots$, the statement has just been proven for $N=M$.
Moreover, any subsequent transformation is now the conventional Darboux
transformation.
Therefore,
the proof in this case follows the same lines as in
\cite{pecheritsin:04}.
\end{proof}

\subsection{Transformation of potential}
Now we will show how to calculate the matrix potential of
the \Sc equation obtained after $N$ Darboux transformations
with $M(<N)$ singular transformations.
 Writing the transformed potential in the form
\[
V_N = V_0 + 2 \p F_N
\]
where
\begin{equation} \label{F_N-def}
F_N = F_{N-1} + (\p Y_N) Y_N^{-1}, \quad F_0 = 0, \quad Y_N =
L_{(N-1) \leftarrow 0} \cU_N\,,
\end{equation}
we emphasis a recursive character of the procedure.

\begin{teo}\label{teo:mShF_N}
Let the matrix $F_N$ be defined by the recursion
 (\ref{F_N-def}).
 Then the elements $f^N_{i,j}$ of the matrix $F_N$
 are expressed in terms of transformation matrices $\cU_k$,
 $k=1,\ldots ,N$
 as follows
\[
f^N_{ij} = \frac{|W_{i,j}(\cU_1, \ldots, \cU_M; \cU_{M+1}, \ldots,
\cU_N)|} {|W(\cU_1, \ldots, \cU_M; \cU_{M+1}, \ldots, \cU_N)|}\,.
\]
 where $W(\cU_1, \ldots, \cU_N)$ is defined by $(\ref{W-def})$
 and $W_{ij}(\cU_1, \ldots, \cU_N)$ is given in $(\ref{Wij-def})$.
\end{teo}

We do not dwell on the proof of the theorem
since using Lemma~\ref{phij-dif} makes applicable with short modifications
 the existing proof of the similar
theorem \cite{pecheritsin:04}.

\section{Determinant representation of the
Kohlhoff-von Geramb
solution of the Marchenko equation}

Note that the theorems from the previous section are not related to a
spectral problem.
They permit to construct matrix-valued potentials for
which the multi-component \Sc equation,
 as a set of second order
differential equations of a special type,
 can be solved exactly.
In this section, as an illustration of this approach, we will apply
obtained formulas for
constructing the simplest local $2\times2$
 potential matrix describing the $^3S_1$ $-$ $^3D_1$
 neutron-proton elastic scattering.
Thus we will give a determinant representation of the potential
previously obtained by Kohlhoff and von Geramb \cite{kohlhoff:93} who used
Marchenko \cite{marchenko:55} inversion method.

Note that in this case $x$ is the radial variable and we will use the
conventional notation for it, $x\equiv r\in(0,\infty)$.

We start with a diagonal matrix potential
\[
V_0=\left(
   \begin{array}{cc}
   0 & 0 \\
   0 & 6/r^2
  \end{array}
   \right), \qquad r\in (0,\infty)
\]
describing the non-interacting free particle in $s$ (subsystem (I))
and $d$ (subsystem (II)) states.
Thus we have here $m=1$ and $n=2$.

Jost
\[
f_s(kr)={\rm e}^{ikr}\,,\qquad f_d(kr)=
{\rm e}^{ikr}\left(1+\frac{3i}{kr}-\frac{3}{(kr)^2}\right),
\]
and regular
\[
\fl
\varphi_s(kr)=i\sin(kr)\,,\qquad
\varphi_d(kr)=\frac{i \left[ (3-k^2 r^2) \sin(kr) - 3 k r
\cos(kr) \right]}{ k^2 r^2}\,.
\]
one-channel $s$ and $d$ partial wave solutions are well known.

We will now realize
a four fold transformation over the composite system (I)+(II)
choosing first two transformations singular with the
transformation matrices $\cU_1$, $\cU_2$ having
 the structure (\ref{cU})
\[
\cU_1=\left(
   \begin{array}{cc}
\varphi_s(ik_1r) & 0 \\
   0           & 1
  \end{array}
   \right),\qquad
   \cU_2=\left(
   \begin{array}{cc}
   f_s(-ik_2r) & 0 \\
   0           & 1
  \end{array}
   \right)
\]
and two other transformations regular with the transformation matrices
\begin{equation}\label{cu34}
\cU_3=\left(
   \begin{array}{cc}
   -\varphi_s(\kappa r)N(-\kappa)  & i f_s(\kappa r)N(\kappa) \\
   -i\varphi_d(\kappa r)             & f_d(\kappa r)
  \end{array}
   \right),\qquad
   \cU_4=\cU_3^*\,.
\end{equation}
Here $N(k)=[(k_1-i k)(k_2-i k)]^{-1}$, $\kappa=\chi(1+i)$, and
$\chi$, $k_1$, $k_2$ are free parameters.
The factors $N(\pm k)$ are introduced in \rf{cu34}
 to guaranty the symmetry, and hence,
Hermitian character of the transformed potential.
Evidently, if every intermediate potential emerging from any step of
transformations is symmetric, the final potential is also symmetric.
 As it is
shown in \cite{samsonov:07}, the matrix of the transformed
potential remains to be symmetric after a first order transformation
provided the self-Wronskian of
the transformation matrix $Y_j$ defined as
${\rm W}[Y_j,Y_j]=Y_j^t Y_j'-(Y_j')^tY_j$ vanishes. The factors
 $N(\pm k)$ are just chosen such that
 ${\rm W}[Y_3,Y_3]={\rm W}[Y_4,Y_4]=0$.

 Note since we are not interested in intermediate potentials, the
 condition that they all are symmetric may be redundant.
 Actually, this is
 sufficient to impose a condition that the resulting action of the chain
 gives a symmetric potential. At present this problem is still waiting for its
 solution.

Having fixed the set of matruces $\cU_j$,
the potential matrix is constructed using Theorem 2,
\be\label{V4}
V_4(r)\equiv V(r)=V_0(r)+\Delta V(r)\,,\quad\Delta V=(\Delta V_{i,j})
\ee
with
\be\label{Dij}
\Delta V_{i,j}=-2\left(\frac{|W_{i,j}(\cU_1,\cU_2;\cU_3,\cU_4)|}%
{|W(\cU_1,\cU_2;\cU_3,\cU_4)|}\right)'
\ee
where matrices $W_{i,j}$ and $W$ are given in (\ref{W-def})
and (\ref{Wij-def}) respectively.

Using the approach developed in \cite{pupasov:10}
one can easily find the $S$-matrix for the potential $V$ \rf{V4},
\rf{Dij}
\[
\fl
S(k)= \frac{1}{k^4+4\chi^4}\left(
\begin{array}{cc}
2\chi^2  & k^2
\\ -k^2 & 2\chi^2
\end{array}
\right) \left(
\begin{array}{cc}
 \frac{(k+i \kappa_1)(k+i \kappa_2)}{(k-i
\kappa_1)(k-i \kappa_2)} & 0
\\ 0  & 1
 \end{array}
 \right)\left(
\begin{array}{cc}
2\chi^2 & -k^2
\\ k^2 & 2\chi^2
\end{array}
\right).
\]
This is just the same $S$-matrix that was used by Kohlhoff and von Geramb for
constructing the simplest potential describing
 neutron-proton elastic collisions by the Marchenko inversion method.
Thus we can state that the obtained potential $V(r)$ \rf{V4}, \rf{Dij}
is
a determinant representation of the Kohlhoff-von Geramb potential.

\begin{figure}[th]
\begin{center}
\epsfig{file=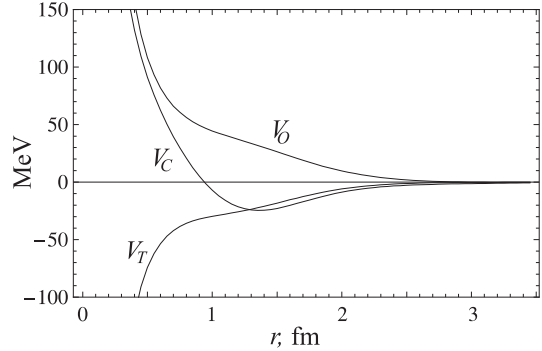, width=7cm}
 \caption{\small
  Exactly solvable potential (central $V_C$, tensor $V_T$ and spin-orbit $V_O$ terms)
 with parameters $k_1=0.944$, $k_2=0.232$~fm$^{-1}$  and $\chi=1.22$~fm$^{-1}$
 taken from \cite{newton:57}.
\label{figSD-hoh}}
\end{center}
\end{figure}

We have to emphasis that according to property established in
 \cite{pupasov:10}, EPP SUSY transformations change the order of channels.
Therefore $V_{1,1}$ and $V_{2,2}$ correspond to $d$-
and $s$-waves respectively.
Also from \rf{V4} and \rf{Dij} we
extract central, $V_C$, tensor, $V_T$, and spin-orbital, $V_O$,
components of the potential
\begin{equation}
\fl
V_C=V_{2,2}\,,\quad V_T=V_{1,2}/\sqrt{8}\,,\quad
V_O=(V_{2,2}-V_{1,2}/\sqrt{2}-V_{1,1}+6/r^2)/3\,.
\end{equation}
Choosing the same parameters $k_1$, $k_2$ and $\chi$
as in \cite{kohlhoff:93,newton:57},
 we
show these potential curves in Figure \ref{figSD-hoh}.

Comparing our results with those shown in figure 13 of paper
\cite{kohlhoff:93}, one can see that in both cases the curves have the same
shape.
Note that the potentials \rf{V4}, \rf{Dij}, as well as
potentials constructed in \cite{kohlhoff:93},
are singular at the origin
and by this reason
they differ essentially
 from potentials obtained by
Newton and Fulton  \cite{fulton:55,newton:57}.
We believe that the difference with Ref. \cite{newton:57}
may be explained by a somewhat different input parameters.
It is well-known
that there is a family of potentials with the same scattering matrix,
when there are bound states.
In our case there is a single bound state and, hence,
there could exist
 two additional parameters,
$A_d$ and $A_s$, known as the asymptotic normalization constants
 which provide an iso-phase deformation
of the potential with the given scattering matrix.
The values of
these free parameters are fixed in both our model and the approach of
 Kohlhoff-von Geramb by the following reason.
Physically acceptable components of a matrix potential should not have
long-range tails.
By this reason, corresponding asymptotic normalization
constants can be determined  directly from the $S$-matrix residue at the
bound state pole \cite{stoks:88}.
 In particular, the ratio of the asymptotic normalization
constants reads
\[
\eta=\frac{A_d}{A_s}=
\frac{{\rm res}S_{2,1}(k=i k_2)}{{\rm res}S_{1,1}(k=i k_2)}\,.
\]
For the potential $V(r)$  \rf{V4}, \rf{Dij},
as well as for the Kohlhoff-von Geramb potential,
the parameter of $\eta$ is fixed by the value
 $\eta=k_2^2/(2\chi^2)=0.018081$.

\section{Conclusion}

As it was pointed out by Andrianov et al \cite{andrianov:97},
an essential feature of matrix Darboux transformations is that the
coefficient before the derivative may be a singular matrix.
 We call such transformations {\it singular transformations}.
 Often these transformations allow one to find hidden symmetries of
the problem \cite{andrianov:97}.
Moreover, any one-channel transformation,
being considered from a multichannel viewpoint, becomes singular.
This leads to a problem of incorporating such transformations
into a chain of conventional (i.e. regular) transformations.

 For the resulting action of the chain of
 conventional matrix Darboux transformations,
 there exists a generalization  \cite{pecheritsin:04} of the well known for
 the one-channel case Crum-Krein formulas \cite{Crum,Krein}.
 In the present paper, a method is developed which allows one to use
one-channel and coupled-channel SUSY transformations on equal terms.
We
 reformulated results obtained in \cite{pecheritsin:04} such that singular
 transformations of a special type are included into the chain as links
 of the same rights as the conventional (regular) transformations
 except that they should be realized before the regular transformations.
This approach together with recently introduced eigenphase preserving
 transformations \cite{pupasov:10}, may become an alternative to the
 Gelfand-Levitan-Marchenko method of solving the inverse scattering
 problem with the advantage that no needs exist for solving any integral equation.
 The values of the $S$-matrix poles are incorporated into the the potential as
 parameters.

We would like to emphasis that the use of exponential and spherical Bessel
functions usually permits to fit the experimental scattering data with a
very high precision.
This means that both the potential and solutions of
the corresponding \Sc equation are expressed via determinants
(see theorems 1 and 2 above)
containing elementary functions only.
 We believe that this is a big
advantage as compared to the Gelfand-Levitan-Marchenko method.
Moreover, in contrast to the usual scheme where one applies
transformations step by step, our approach permits one to skip all
intermediate calculations and obtain the final potential directly in terms
of solutions of the initial \Sc equation,
i.e. usually in terms of elementary functions.

 As an illustration of the method,
 using a special six-pole representation of the $S$ matrix,
  we derived a determinant representation of
 the Kohlhoff-von Geramb \cite{kohlhoff:93}
  solution to the Marchenko equation.

A drawback of the method, that we see, is that no a general recipe
is known for
choosing transformation matrices in a way to
produce an Hermitian potential.
Yet, in any concrete calculation this is enough
 to chose transformation matrices such that any intermediate potential is
 Hermitian. This is possible since corresponding condition is known
 \cite{samsonov:07} but the general problem is waiting for its solution.

   \ack
The work is partially supported
by the  Russian government
 under contracts 02.740.11.0238, P1337 and P2596.
AMP and BFS are very indebted to D. Baye and J.-M. Sparenberg for
 numerous stimulating discussions
 about possible applications of SUSY transformations in nuclear physics.

 \setcounter{equation}{0}
 \renewcommand{\theequation}{A.\arabic{equation}}

 \section*{Appendix A}

 Here we formulate the Sylvester identity \cite{Gantmaher}.
 Consider a square matrix of dimension $p+q$, $p,q=1,2,\ldots $
 \be
  A=\left(
  \begin{array}{cccccc}
  a_{1,1} & \ldots & a_{1,p} & a_{1,p+1} & \ldots & a_{1,p+q}\\
         &  \ldots &       &          & \ldots &         \\
  a_{p,1} & \ldots & a_{p,p} & a_{p,p+1} & \ldots & a_{p,p+q}\\
  b_{1,1} & \ldots & b_{1,p} & b_{1,p+1} & \ldots & b_{1,p+q}\\
         &  \ldots &       &          & \ldots &         \\
  b_{q,1} & \ldots & b_{q,p} & b_{q,p+1} & \ldots & b_{q,p+q}\\
  \end{array}
  \right).
 \ee
 Let $a$ be the submatrix of dimension $p\times p$ composed of the
 elements $a_{i,j}$, $i,j=1,\ldots ,p$. If to the bottom of $a$
 we add a line of elements $b_{k,1}$, \ldots , $b_{k,p}$, to
 the right of $a$ we add a column of elements $a_{1,p+l}$, \ldots , $a_{p,p+l}$
 and the right bottom corner we fill with the element $b_{k,p+l}$,
 we obtain a square matrix $m_{j,l}$.
 One says that   $m_{j,l}$ is
 obtained from $A$ by {\em embordering} the block $a$ with $k$th row and
 $(p+l)$th column. The determinant $| m_{j,l} |$ is called an
 {\em embordering  minor}
  in the determinant $|A|$. Since $k$ and $l$ can take the values
 $k,l=1,\ldots ,q$ one has $q\times q$ embordering minors from
 which one can construct the matrix $M=(m_{j,l})$. The Sylvester
 identity relates the determinants $|M|$, $|A|$ and $|a|$ as
 follows:
 \be
  |M|=|a|^{q-1}\,|A|\,.
 \ee

\end{document}